\documentclass[twocolumn,preprintnumbers,amsmath,amssymb,notitlepage,nofootinbib,longbibliography,superscriptaddress,aps,pra,10pt]{revtex4-1}

	\usepackage{natbib}
	\usepackage{graphicx}
	\usepackage{dcolumn}
	\usepackage{dsfont}
	\usepackage{bm}
	\usepackage{amsmath}
	\usepackage{complexity}
	\usepackage{times}
	\usepackage{amsmath}
	\usepackage[usenames, dvipsnames]{color}
	\usepackage{url}
	\usepackage[colorlinks=true,breaklinks=true,allcolors=blue]{hyperref}
	\usepackage{microtype}

	\newtheorem{problem}{Problem}
	\newcommand{\defeq}{\mathrel{\mathop:}=}

\begin{document}

	\title{Reinforcement Learning Decoders for Fault-Tolerant Quantum Computation}

	\author{Ryan Sweke}
	\affiliation{Dahlem Center for Complex Quantum Systems, Freie Universit\"{a}t Berlin, 14195 Berlin, Germany}
	\author{Markus S. Kesselring}
	\affiliation{Dahlem Center for Complex Quantum Systems, Freie Universit\"{a}t Berlin, 14195 Berlin, Germany}
	\author{Evert P.~L. van Nieuwenburg}
	\affiliation{Institute for Quantum Information and Matter, Caltech, Pasadena, CA 91125, USA}
	\author{Jens Eisert}
	\affiliation{Dahlem Center for Complex Quantum Systems, Freie Universit\"{a}t Berlin, 14195 Berlin, Germany}
	\affiliation{Department of Mathematics and Computer Science, Freie Universit\"{a}t Berlin, 14195 Berlin}

	\date{\today}

	\begin{abstract}
		Topological error correcting codes, and particularly the surface code, currently provide the most feasible 
		roadmap towards large-scale fault-tolerant quantum computation.
		As such, obtaining fast and flexible decoding algorithms for these codes, within the experimentally relevant context of faulty syndrome measurements, is of critical importance.
		In this work, we show that the problem of decoding such codes, in the full fault-tolerant setting, can be naturally reformulated as a process of repeated interactions between a decoding agent and a code environment, to which the machinery of reinforcement learning can be applied to obtain decoding agents.
		As a demonstration, by using deepQ learning, we obtain fast decoding agents for the surface code, for a variety of noise-models.
	\end{abstract}

\maketitle

\section{Introduction}\label{s:introduction}
	In order to implement large scale quantum computations it is necessary to be able to store and manipulate quantum information in a manner that is robust to the unavoidable errors introduced through interaction of the physical qubits with a noisy environment.
	The known strategy for achieving such robustness is to encode a single logical qubit into the state of many physical qubits, via a quantum error correcting code, from which it is possible to actively diagnose and correct errors that may occur~\cite{Terhal15,Campbell17}.
	While many quantum error correcting codes exist, topological quantum codes~\cite{Kitaev03, Dennis02, Preskill17lectures, Nayak08, Pachos12, Terhal15, Brown16, Campbell17}, in which only local operations are required to diagnose and correct errors, are of particular interest as a result of their experimental feasibility~\cite{Reed12, Barends14, Nigg14, Corcoles15, Albrecht16, Takita16, Linke17}.
	In particular, the surface code has emerged as an especially promising candidate for large-scale fault-tolerant quantum computation, due to the combination of its comparatively low overhead and locality requirements, coupled with the availability of convenient strategies for the implementation of all required logical gates~\cite{Fowler18,Litinski18b}.
	In fact, current road maps towards the realization of robust quantum computing have identified surface code based approaches as the most feasible methodology for achieving this goal~\cite{Roadmap}.

	However, the known realistic topological quantum error correcting codes, including the surface code, are not self-correcting, and are therefore not robust to natural thermal noise.
	For this reason one has to actively diagnose and correct for errors, and as such, in any code-based strategy for fault-tolerant quantum computation decoding algorithms play a critical role.
	At a high level, these algorithms take as input the outcomes of syndrome measurements (which provide a diagnosis of errors that have occurred on the physical qubits), and provide as output a suggestion of corrections for any errors that may have occurred during the computation.
	In practice, these decoding algorithms have to be extremely fast - in particular, one has to be able to decode faster than the rate at which errors occur.
	As such, the development of decoding algorithms constitutes a serious bottleneck in the realization of fault-tolerant quantum computers and are key to gaining an understanding of quantum computing in realistic regimes.

	It is particularly important to note that in any physically realistic setting, the required syndrome measurements are obtained via small quantum circuits, and are therefore also generically faulty.
	For this reason, while the setting of perfect syndrome measurements provides a paradigmatic test-bed for the development of decoding algorithms, any decoding algorithm which aims to be experimentally useful 
	must necessarily be capable of dealing with such faulty syndrome measurements.
	Additionally, such algorithms should also be capable of dealing with experimentally relevant noise models, as well as be fast enough to not present a bottleneck to the execution of computations, even as the size of the system (i.e.\ the code distance) 
	grows.

	As of yet, it is precisely the development of decoders applicable to the fault-tolerant setting that constitutes a particular challenge.
	However, due to the importance of decoding algorithms for fault-tolerant quantum computation, several approaches have been developed, each of which tries to satisfy as many of the experimentally required criteria as possible.
	Perhaps most prominent are algorithms based on minimum-weight perfect matching subroutines~\cite{Fowler13}, although alternative approaches based on techniques such as the renormalization group~\cite{Duclos2010} and locally operating cellular automata~\cite{Herold15,herold2017cellular,Kubica2018} have also been put forward.
	These algorithms solve the problem in principle, but may well be too slow in realistic settings.

	Recently, techniques from machine learning have begun to find application in diverse areas of quantum physics - such as in the efficient representation of many-body quantum states~\cite{WFcarleo2017solving,WFcarleo2018constructing,WFgao2017efficient}, the identification of phase transitions~\cite{PTvan2017learning,PTPhysRevB.97.134109,PTcarrasquilla2017machine,broecker2017quantum,PhysRevB.95.245134}, and the autonomous design of novel experimental set-ups~\cite{melnikov2018active,fosel2018reinforcement} - and in an attempt to tackle the issue of fast decoding various neural-network based decoders have also been proposed~\cite{Torlai10, Varsamopoulos17, Krastanov17, chamberland2018deep, Baireuther18a, Baireuther18b, Breuckmann18,Ni18}.
	In particular, previously proposed neural network decoders promise extremely fast decoding times~\cite{chamberland2018deep}, flexibility with respect to the underlying code and noise model~\cite{chamberland2018deep,Baireuther18a,Baireuther18b,Breuckmann18} and the potential to scale to large code distances~\cite{Ni18, Breuckmann18}.
	However, despite this diversity of proposed decoding algorithms, and the clear potential of machine learning based approaches, there is as of yet no algorithm or technique which clearly satisfies all the required experimental criteria listed above.
	As such, there remains room for improvement and new techniques, particularly within the fault-tolerant setting.

	Simultaneously, the last few years have also seen impressive advances in the development of deep reinforcement learning algorithms, which have allowed for the training of neural network based agents capable of obtaining super-human performance in challenging domains such as Atari~\cite{RLMnih15,RLschaul2015prioritized,RLvan2016deep,RLwang2015dueling}, Chess~\cite{RLsilver2017mastering} and Go~\cite{RLSilver2016,RLSilver17b}.
	These techniques are particularly powerful in situations where it is necessary to learn strategies for complex sequential decision making, involving consideration of the future effects of ones actions.
	The decoding problem within the context of fault-tolerant quantum computation is precisely such a problem.
	Given the significant challenges involved in the development of fast decoders for the fully fault-tolerant setting and the clear parallels between decoding and the domains in which reinforcement learning techniques have excelled, it is natural to ask both the extent to which these techniques can be applied to the decoding problem and the advantages that such an approach would offer over alternative methods.

	In this work we introduce a novel reinforcement learning based framework for obtaining a new class of decoding algorithms applicable to the setting of fully fault-tolerant quantum computation.
	Importantly, we discuss the advantages of such an approach over the previously proposed neural network decoders, and argue that the framework presented here both lays the foundations and provides a toolbox for obtaining decoders satisfying all the required criteria.
	In particular, within the newly established framework to be presented here, we interpret the decoder as an agent performing discrete, sequential actions on an environment which is defined by a quantum error correction code.
	This conceptual framework allows for the application of various deep reinforcement learning algorithms to obtain neural network based decoding agents.
	As a demonstration, we then utilize deepQ learning to obtain fast surface code decoders, for a variety of noise models, in the fully fault-tolerant setting.
	These results provide a foundation for extension via both more sophisticated reinforcement learning techniques and neural network models.

	In this work we begin by providing an introductory overview of the surface code in Section~\ref{s:the_surface_code}, before presenting a description of the decoding problem for fault-tolerant quantum computation in Section~\ref{s:the_decoding_problem}.
	After a brief introduction to the formalism of reinforcement learning and $Q$-functions in Section~\ref{s:reinforcement_learning} we are then able to provide the conceptual re-framing of decoding as a reinforcement learning problem in Section~\ref{s:decoding_as_rl}, representing one of the primary results of this work.
	In Section~\ref{s:results} we then present deepQ surface code decoders for a variety of noise models, before finally in Section~\ref{s:conclusions} we discuss both the advantages and disadvantages of the approach presented here, along with various potential strategies for building upon the results presented in this work.

\section{The Surface Code}\label{s:the_surface_code}

	We begin by providing a brief description of the surface code.
	The framework and methods presented in this work are not restricted to the surface code however, and may be applied to any stabilizer code.
	The restriction to the surface code is made both for simplicity of presentation and experimental relevance.
	We will focus on presenting the essential elements of the surface code and refer to more complete treatments for details~\cite{Gottesman97, Terhal15,Litinski18b}.

\begin{figure}
		\centering
		\includegraphics[width=0.8\linewidth]{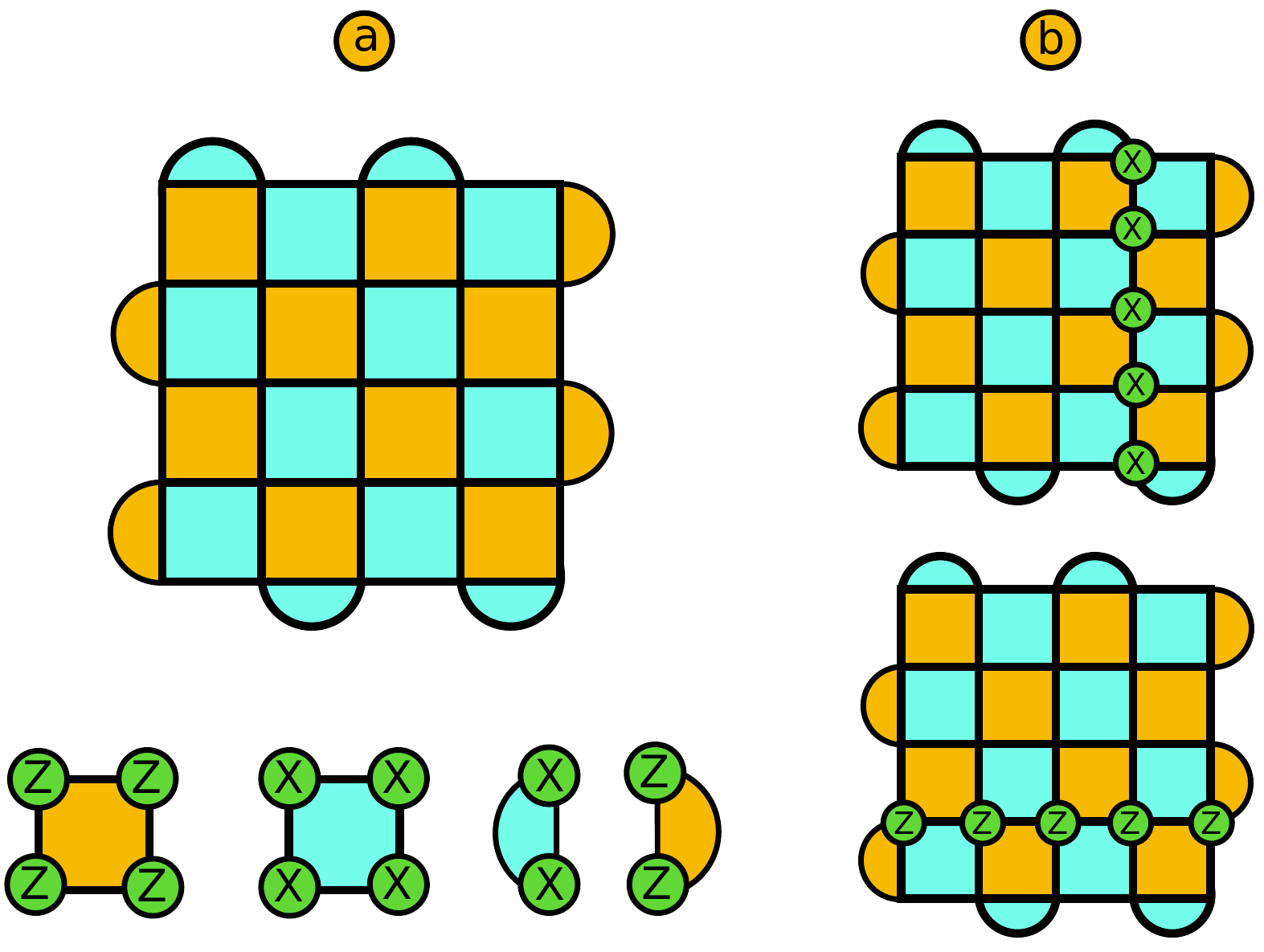}
		\caption{
			An overview of the $5 \times 5$ surface code.
			(a) We consider square lattices, with a physical data qubit on each vertex.
			The colored plaquettes indicate stabilizer operators as defined in Eq.~\eqref{e:stabilizer_definition}.
			(b) Logical $X_L$ and $Z_L$ operators for the surface code are given by continuous strings of single qubit $X$ or $Z$ operators connecting the top and bottom or left and right boundaries of the code respectively.
		}
		\label{f:surface_code}
	\end{figure}

	We will consider $d\times d$ lattices with a \textit{physical} data qubit on each vertex $v$, as illustrated in Fig.~\ref{f:surface_code} for $d=5$.
	The collective state of all qubits on the lattice is an element of the Hilbert space $\mathcal{H} = \mathbb{C}^{2^{(d\times d)}}$.
	We associate stabilizer operators with each colored plaquette of the lattice.
	Stabilizers on blue (orange) plaquettes are operators which apply Pauli $X$ ($Z$) flips to all qubits on the vertices of the plaquette.
	Specifically, denoting the set of all blue (orange) plaquettes as $B_p$ ($O_p$) we define the stabilizer $S_p$ on plaquette $p$ as,
	\begin{equation}\label{e:stabilizer_definition}
		S_p = \bigotimes_{v\in p} \sigma_v \quad \text{where }
		\begin{cases}
			\sigma_v = X_v \quad \text{if } p \in B_p,\\
			\sigma_v = Z_v \quad \text{if } p \in O_p.
		\end{cases}
	\end{equation}
	All stabilizers are mutually commuting and have eigenvalues $\pm 1$.
	This allows for the introduction of a fictitious Hamiltonian $H = -\sum_p S_p$ from which the surface code $\mathcal{H}_\mathrm{sc} \subset \mathcal{H}$ is defined to be the ground state space of $H$.
	Alternatively, this space consists of all simultaneous $+1$ eigenstates of all stabilizers.
	This subspace is two dimensional, i.e.\ $\mathcal{H}_\mathrm{sc} \simeq \mathbb{C}^2$, and hence can encode a single \textit{logical} qubit.
	Logical operators are operators which preserve the code space, and can therefore be used to manipulate the state of the logical qubit.
	Fig.~\ref{f:surface_code} shows logical $X$ ($Z$) operators, denoted $X_L$ ($Z_L$), which are continuous strings of single vertex $X$ ($Z$) operators connecting the top and bottom (left and right) boundaries of the lattice.

	To illustrate the motivation behind such an encoding, let us examine the consequences of a single qubit Pauli flip on a physical data qubit.
	If we assume that the initial state vector $|\psi\rangle \in \mathcal{H}_\mathrm{sc}$ is an element of the code space, then the subsequent state 
	vector $|\psi'\rangle \not \in \mathcal{H}_\mathrm{sc}$ will no longer be an element of the code space.
	In particular, $|\psi'\rangle$ will be an eigenstate with eigenvalue $-1$ of at least one stabilizer.
	We say that $|\psi'\rangle$ violates these stabilizers, as illustrated by red circles in Fig.~\ref{f:surface_code_examples}~(a).
	The \textit{syndrome} of a state is a list of the outcomes of a simultaneous measurement of all the stabilizers, each of which takes the value $\pm 1$.
	Given that a single Pauli flip occurred on a single physical data qubit, by analyzing the syndrome we may be able to identify and correct this error, in the process conserving the logical qubit state.
	This process of \textit{decoding} is discussed in the next section.

	In the terminology of stabilizer codes and quantum error correction,  such a surface code on a $d\times d$ lattice is a 
	$[[d^2,1,d]]$ code~\cite{Gottesman97}.
	This means that $d^2$ physical qubits data qubits are required to encode a single logical qubit, with code distance $d$.
	Specifically, the \emph{distance of a stabilizer code} is the weight of the minimal-weight non-trivial logical operator - i.e.\ the minimal weight Pauli operator that both preserves the code subspace and acts non-trivially within this code subspace.
	In particular, this quantity characterizes the error correcting capabilities of a given code: if a logical qubit is affected by the action of a Pauli operator, the weight of which is less than half the code distance, then one can successfully recover from the error by applying the minimal weight Pauli operator which returns the altered state to the code subspace.

\section{The Decoding Problem}\label{s:the_decoding_problem}

	With the foundations from the previous section, we can now formulate a simple preliminary version of the decoding problem.

	\begin{problem}[Decoding problem]Assume that at $t=0$ one is given a state vector $|\psi\rangle = \alpha |0_L\rangle + \beta |1_L\rangle \in \mathcal{H}_{\mathrm{sc}}.$ \textit{At some time }$t_1>0$ \textit{a syndrome measurement is performed which indicates that one or more stabilizers are violated - i.e.\ some errors have occurred on physical data qubits.
	From the given syndrome, determine a set of corrections which should be applied to the code lattice such that the subsequent state} $|\psi'\rangle$ \textit{is equal to the initial state} $|\psi\rangle.$ 
	\end{problem}
	
\begin{figure}
		\centering
		\includegraphics[width=1\linewidth]{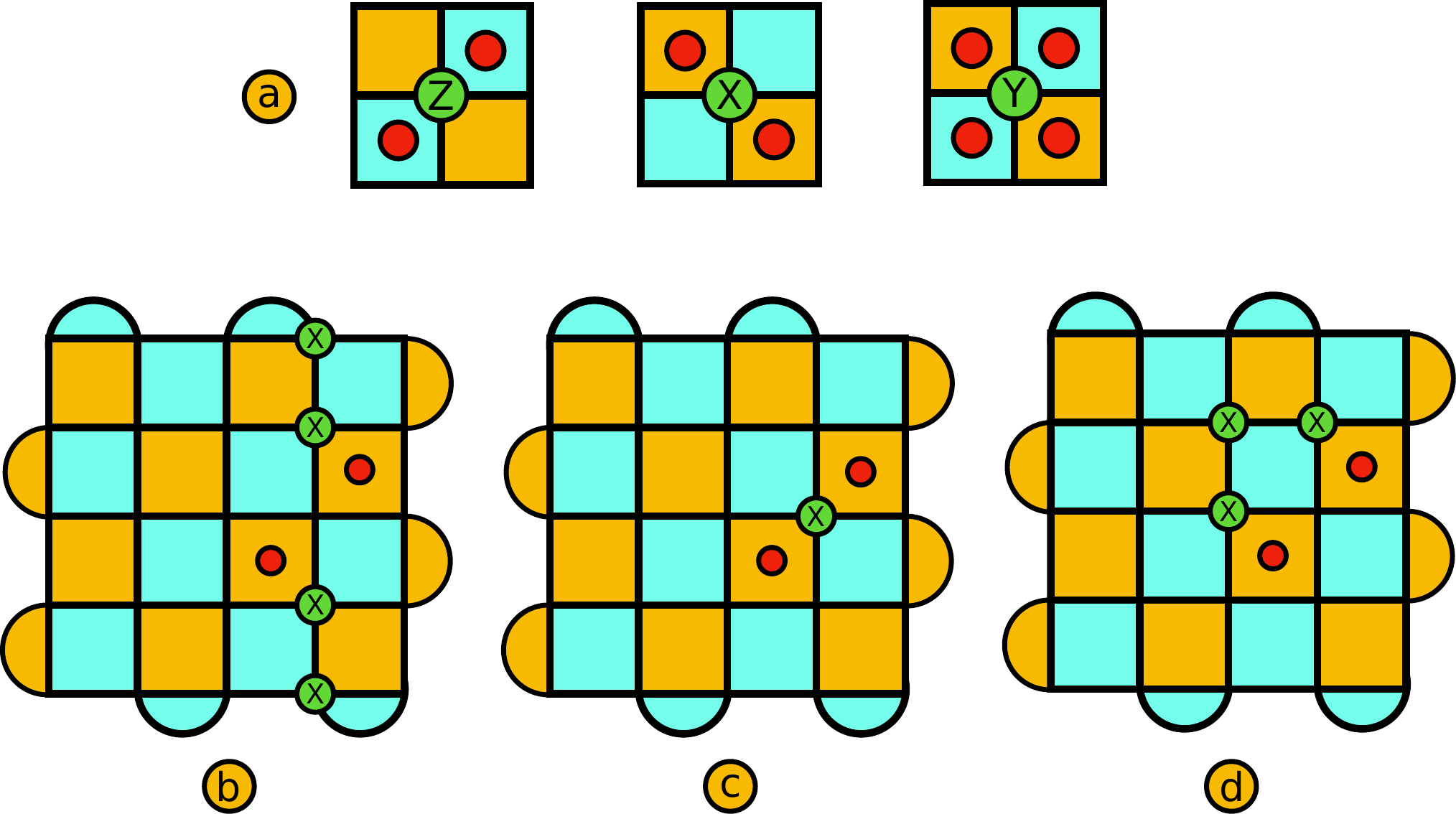}
		\caption{
			(a) Single qubit Pauli flips violate surrounding stabilizers.
			(b-d) Strings of Pauli flips only violate stabilizers at the endpoint of the string.
			Multiple error configurations can give rise to the same syndrome.
			They can differ by stabilizers, as for example in (c) and (d), or by logical operators, see (b) and (c).
		}
		\label{f:surface_code_examples}
	\end{figure}

	\noindent Before proceeding to discuss more subtle and technical versions of the decoding problem, let us examine why even the above problem is indeed potentially difficult.
	The most important observation is that the map from error configurations to syndromes is many-to-one, i.e.\ many different sets of errors can lead to the same syndrome.
	As an example, consider the error configurations illustrated in Fig.~\ref{f:surface_code_examples}~(b-d), all of which lead to the same syndrome.
	If the probability of an error on a single physical data qubit is low, given such a syndrome one might reasonably assume that the error shown in (c) occurred, as one error on a single qubit is a more likely event than errors on multiple qubits.
	Given this reasoning, one might then suggest to correct by applying an $X$ flip on the physical data qubit in the third row and fourth column.
	If indeed the error shown in (c) occurred, the post-correction state would be error-free, thus preserving the initial state.
	However, if the error pattern shown in (d) occurred, this set of errors combined with the applied $X$ flip would implement a stabilizer.
	Since the original state was a simultaneous $+1$ eigenstate of all the stabilizers, and stabilizers act trivially on logical states, the proposed correction indeed preserves the initial logical state.
	Finally, if the error in (b) occurred, then the combination of the original error with the correction would implement the logical $X_L$ operator.
	As a result, even though the post-correction state is back in the code space, it will be in a different logical state.
	Thus, the information we were trying to preserve would have been corrupted.
	From this simple example one can see that most often solving the decoding problem as stated above involves deciding, given 
	an inherently ambiguous syndrome and (possibly imperfect) knowledge of the underlying error model, which error configuration most likely occurred.
	The mathematical structure underlying the decoding problem is that of homology, which provides a 
	concise representation of the relationship between errors and syndromes~\cite{FastDecoders}.
	In this language,
	a decoder determines the approximate relative likelihood of different homology classes -- equivalence classes of 
	error patterns -- of given error configurations that are captured by the syndrome.

	In addition to the inherent difficulty resulting from syndrome ambiguity, in experimental settings the process of extracting the syndrome is 
	its itself subject to noise.
	That is to say, one must reasonably assume that the syndrome itself may be faulty~\cite{tomita2014low,stephens2014fault}.
	In practice, each stabilizer may be associated with a physical ancilla qubit.
	The syndrome value for that particular stabilizer is obtained by first executing a small quantum circuit which entangles the ancilla with each of the physical data qubits on which the corresponding stabilizer is supported.
	The syndrome value is then extracted via a measurement of the ancilla qubit.
	In order to fully account for errors during the process of syndrome extraction one should therefore model this entire circuit, in which errors can occur on both the data qubits and ancilla qubits at each time step.
	Moreover, errors on the ancilla qubits can propagate onto the data qubits via the required entangling gates.

	\begin{figure*}
		\centering
		\includegraphics[width=0.6\textwidth]{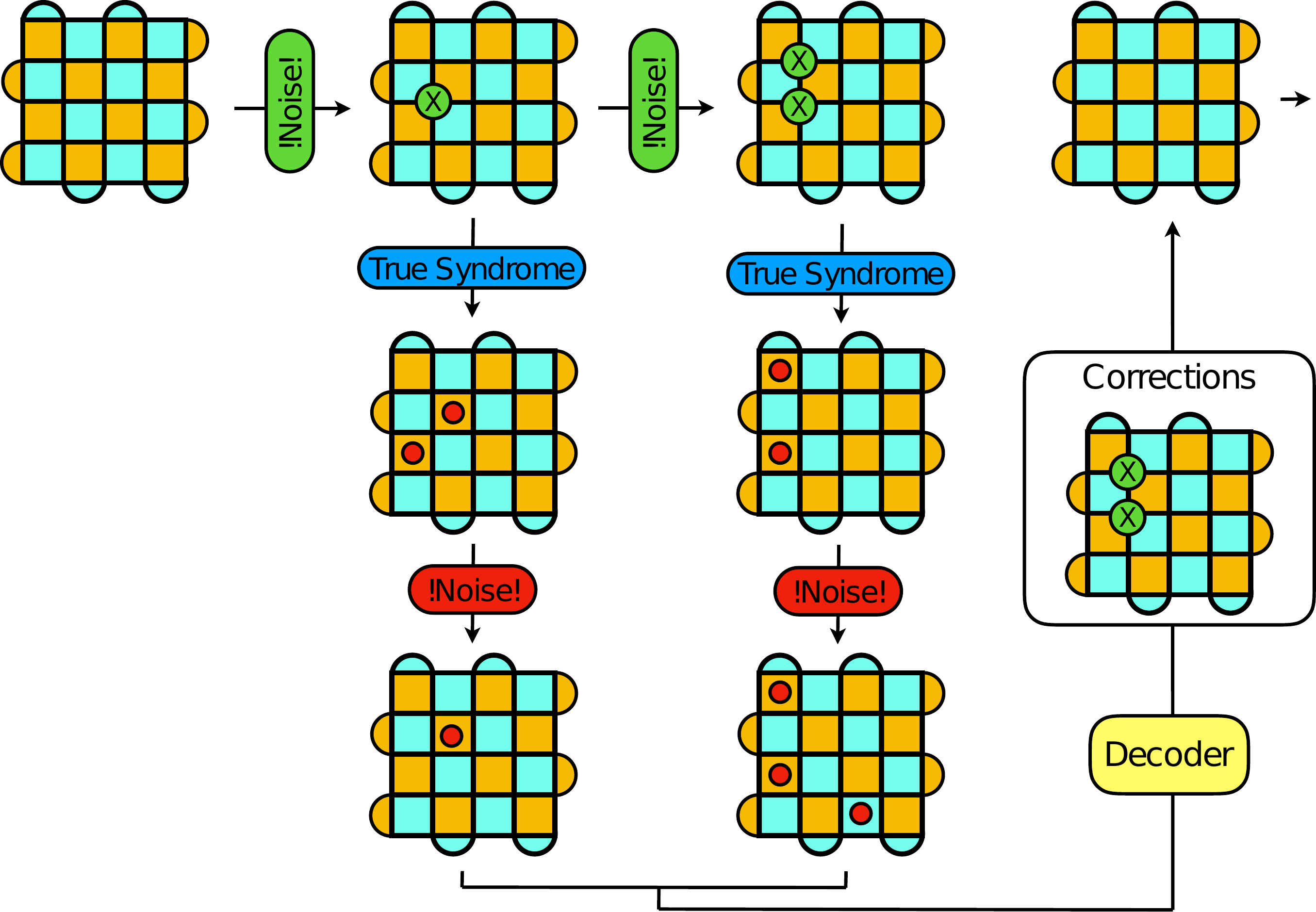}
		\caption{
			A typical decoding cycle is illustrated for the simplified faulty measurements scenario in which one imagines each time step consisting of an initial physical error process generating errors on the data qubits, followed by a second measurement error process which corrupts the true syndrome.
			The decoding algorithm then has access to a sequence of potentially faulty syndromes.
		}
		\label{f:decoding_problem}
	\end{figure*}

	The essential aspect of the additional difficulty from faulty syndrome measurements can be phenomenologically modeled by imagining each time step as consisting of two distinct error processes~\cite{stephens2014fault}, as illustrated in Fig.~\ref{f:decoding_problem}.
	In the first error process, an error occurs on each data qubit with some probability.
	One then imagines extracting the perfect syndrome before a second error process occurs, in which with a given probability an error occurs on each stabilizer measurement outcome.
	In this realistic scenario, lacking only the propagation of errors during syndrome extraction, single syndrome measurements are hence no longer reliable.
	Decoding in the fault-tolerant setting therefore typically requires providing a set of sequential syndrome measurements, and it should be clear at this point that by including the requirement of fault tolerance, the nature of the decoding problem changes substantially.

	Finally, in the context of surface code based fault-tolerant quantum computing, all logical gates are implemented either via protocols which also involve an inherent decoding procedure or do not spread errors.
	To be specific, it is sufficient for universal quantum computing to be able to implement both Clifford and $T$ gates~\cite{Nebe01,Bravyi05}.
	In contemporary proposals for surface code based quantum computing~\cite{Litinski18b,Fowler18}, 
	protocols are known for implementing Clifford gates either by tracking, via code deformation~\cite{Brown17} or via lattice 
	surgery~\cite{Horsman12,Litinski18}.
	Code deformation and lattice surgery requires several decoding cycles.
	Non-Clifford gates, such as the $T$ gate, can be performed fault-tolerantly via gate teleportation using magic states.
	High quality magic states can be obtained via magic state distillation, which requires only Clifford gates and faulty magic states~\cite{Bravyi05}.
	As such the goal of decoding idling logical qubits within a quantum computation should be to suppress errors to the extent that any of the above procedures can succeed with high probability.
	Therefore, we can relax the requirement that the decoding process should return the post-error state to the initial state in the code space.
	In Section~\ref{s:decoding_as_rl} we discuss a proxy criterion for decoding success within this framework.

\section{Reinforcement Learning and q-Functions}\label{s:reinforcement_learning}

\begin{figure}
		\centering
		\includegraphics[width=0.4\textwidth]{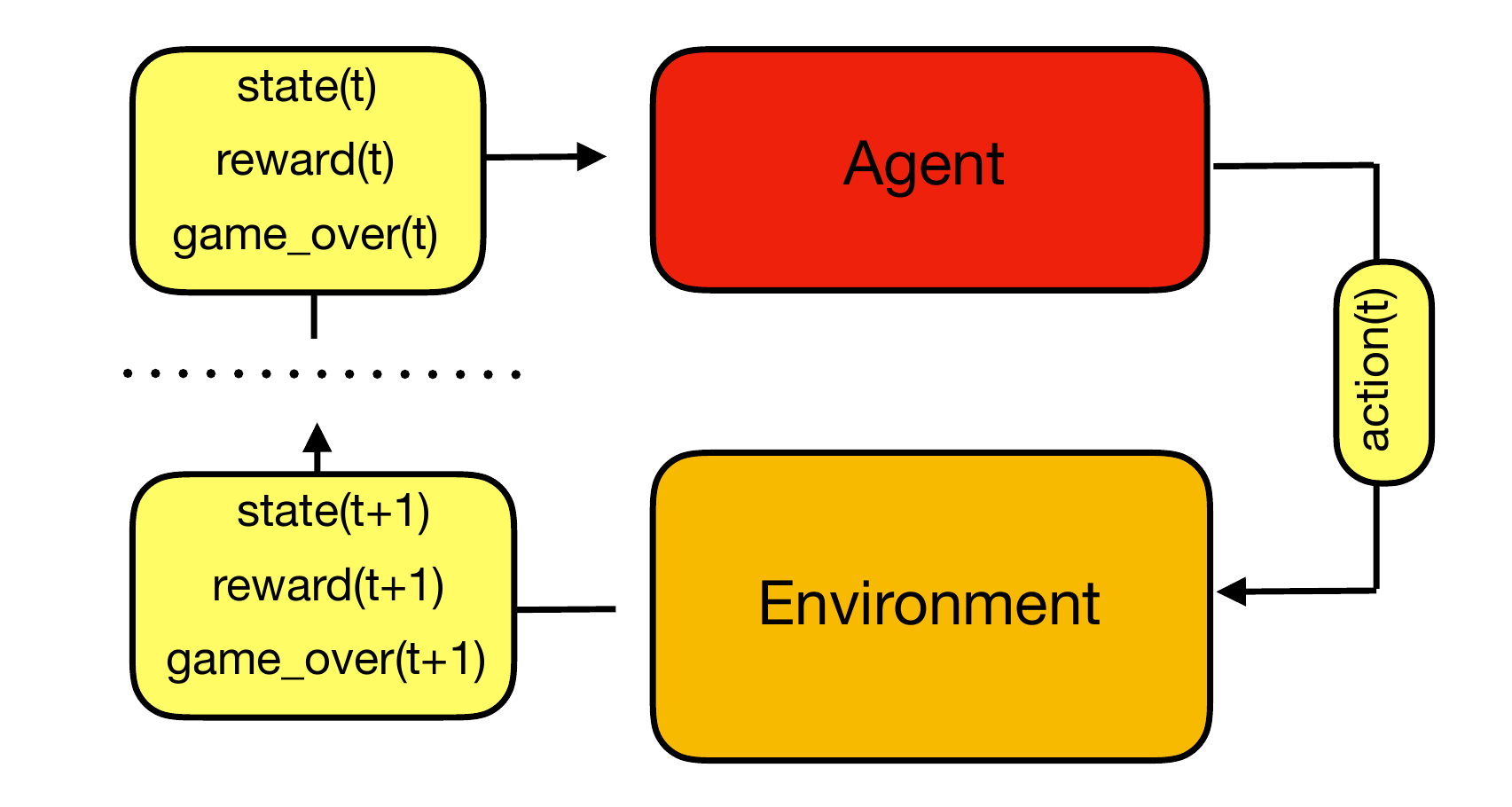}
		\caption{
			An illustration of the signals passed between an agent and the environment through the duration of a sequential turn based episode.
		}
		\label{f:agent_environment}
	\end{figure}

	In this section we shift focus and introduce some of the fundamental concepts of reinforcement learning and $q$-functions, which will be essential to our rephrasing of the decoding problem in Section~\ref{s:decoding_as_rl}.
	Again, we will keep the discussion brief and refer to Ref.~\cite{RLSutton18} for a more complete treatment.
	A generic reinforcement learning problem considers an agent interacting with an environment, as is illustrated in Fig.~\ref{f:agent_environment}.
	The agent can act on and observe parts of the environment, and is tasked with achieving a problem-specific goal by performing a sequence of actions.
	We typically consider discrete problems, in  which at each step time step $t$ the environment can be described by a state $S_t \in \mathcal{S}$, where $\mathcal{S}$ is referred to as the state space.
	Given a state of the environment, the agent can then choose to perform an action $A_t \in \mathcal{A}$, where $\mathcal{A}$ is referred to as the action space.
	As a result of the agents chosen action, the environment then updates accordingly, entering a new state $S_{t+1}$ and providing feedback to the agent on its choice of action in the form of a scalar reward $R_{t+1}$.
	We will restrict ourselves here to episodic environments, for which there exist a set of terminal states $\mathcal{S}_{\mathrm{terminal}} \subset \mathcal{S}$.
	In such episodic settings, in addition to a scalar reward, the agent also receives a Boolean signal $T_{t+1}$, indicating whether $S_{t+1} \in \mathcal{S}_{\mathrm{terminal}} $ - i.e.\ whether or not it is ``game over''.

	In general, the agent's choice of action, the resulting state of the environment and the returned reward can all be stochastic.
	In the case of finite state and action spaces, the environment can then be formalized via a  classical  \emph{finite Markov decision process (FMDP)} governed by the transition probabilities
	\begin{equation}
		p(s',r|s,a) \defeq \mathrm{pr}(S_t = s',R_t = r|S_{t-1} = s, A_{t-1} = a).
	\end{equation}
	To formalize the decision making process of the agent, we define an agent's \textit{policy} $\pi$, in essence the agent's strategy, as a mapping from states to probabilities of specific actions - i.e. $\pi(a|s)$ is the probability that agent chooses $A_t = a$, given that the environment is in state $S_t = s$.
	For FMDP's we then define the \textit{value} of a state $s$ under policy $\pi$ as,
	\begin{equation}
		v_{\pi}(s) = \mathbb{E}_{\pi}[G_t|S_t = s]  = \mathbb{E}_{\pi} \Big[\sum_{k = 0}^{\infty}\gamma^k R_{t+k+1}\Big| S_t = s \Big] 
	\end{equation}
	$\forall S_t \in \mathcal{S}$.
	The term $G_t$ is the discounted return (discounted cumulative reward), with discount factor $0 \leq \gamma \leq 1$, and is the quantity that the agent is tasked with optimizing.
	In episodic settings the infinite sum terminates whenever state $S_{t+k+1}$ is a terminal state - i.e. $S_{t+k+1} \in \mathcal{S}_{\mathrm{terminal}}$.
	We call $v$ the \textit{state-value function}, providing the expected discounted cumulative reward the agent would obtain when following policy $\pi$ from state $s$.
	It is an important conceptual point to note that by using the metric of the discounted cumulative reward the value of any given state depends not only on the immediate reward obtained by following a specific policy from that state, but includes future expected rewards.
	Hence strategies which involve some element of successful future planning may lead to higher state values.
	As such, we see that the value of a state with respect to a given policy reflects accurately the ability of an agent to achieve its long-term goals when following that policy from that state.

	Similarly to the \textit{state-value function}, we can define the \textit{action-value function}  (referred to as the $q$-function) for policy $\pi$ via
	\begin{align}
		q_{\pi}(s,a) &= \mathbb{E}_{\pi}[G_t|S_t = s, A_t = a]  \nonumber\\
		& = \mathbb{E}_{\pi} \Big[\sum_{k = 0}^{\infty}\gamma^k R_{t+k+1}\Big| S_t = s, A_t = a \Big] .
	\end{align}
	Clearly, the $q$-function with respect to a given policy is conceptually similar to the state-value function, differing only in that it provides the value for state-action pairs.
	Importantly, value functions allow us to place an order over policies, i.e. $\pi > \pi' \iff v_{\pi}(s) > v_{\pi'}(s)\quad \forall s \in \mathcal{S} $.
	This in turn allows us to define an optimal policy $\pi^*$, for which
	\begin{equation}\label{e:fixed_point}
		q_*(s,a) = \mathbb{E}\big[R_{t+1} + \gamma\max_{a'}q_{*}(S_{t+1},a')\big|S_t = s, A_t = a \big].
	\end{equation}
	Note that given the optimal $q$-function it is easy to obtain the optimal strategy. In a given state $s$ simply choose the action $a = \mathrm{argmax}_{a'}[q_*(s,a')]$.

	Given this framework, there are many different approaches and methodologies that can be used to learn optimal policies.
	We will focus here on $q$-learning, the goal of which is to find or approximate $q_*(s,a)$.
	This is generically done via iterative $q$-learning, in which the agent starts with an arbitrary $q$-function and then iteratively refines it on the basis of experience gained from interaction with the environment.
	In particular, in order to generate such experience, the agent uses a policy derived from its $q$-function (possibly in addition to other explorative policies) to choose actions.
	This $q$-function is then periodically updated using Eq.~\eqref{e:fixed_point}, for which $q_*(s,a)$ is a stationary solution~\cite{RLSutton18}.

	The above describes concisely the elements of $q$-learning, however it does not address the general impracticality of storing this $q$-function.
	In most real world applications, the number of valid state-action pairs can be impractically large (e.g. consider the number of possible chess configurations).
	It is precisely to address this problem, and in effect to render $q$-learning applicable in practice, that \textit{deepQ} learning was introduced~\cite{RLMnih15,RLvan2016deep,RLschaul2015prioritized}.
	In particular, in deepQ learning we parameterize $q$ by a neural network, and use Eq.~\eqref{e:fixed_point} to construct the cost function from which the network weights can be updated via a stochastic gradient descent method.
	The learning of the $q$-function is hence done by training a neural network in an online supervised manner, the training instances for which are generated by having the agent explore the environment.
	Specifically, we let the agent interact with the environment via an $\epsilon$-greedy exploration policy, in the process generating experience-tuples of the form
	\begin{equation}
		[S_t,A_t,R_{t+1},S_{t+1},T_{t+1}].
	\end{equation}
	The set of these tuples then provides an experience memory, from which we can periodically sample a training-batch.
	Given such a batch of training instances, the $q$-network is then updated via the cost function
	\begin{align} 
		C &= y_{\mathrm{pred}} - y_{\mathrm{true}}\\
		&= q(S_t,A_t) - \big[R_{t+1} + \gamma\max_{a'}q(S_{t+1},a') \big],
	\end{align}
	which, by comparison with Eq.~\eqref{e:fixed_point}, will be minimized by the optimal policy $q_*$.
	Unfortunately, despite the simplicity of this idea, in practice a variety of tricks - such as separate active and target $q$-networks, double-$q$ learning and dueling networks - are required to achieve stable $q$-learning.
	We refer to the relevant references~\cite{RLMnih15,RLvan2016deep,RLschaul2015prioritized,RLwang2015dueling}, or to the associated code repository~\cite{DeepQDecoding}, for details.

\section{Decoding as a Reinforcement Learning Problem}\label{s:decoding_as_rl}

	\begin{figure*}
		\centering
		\includegraphics[width=0.65\textwidth]{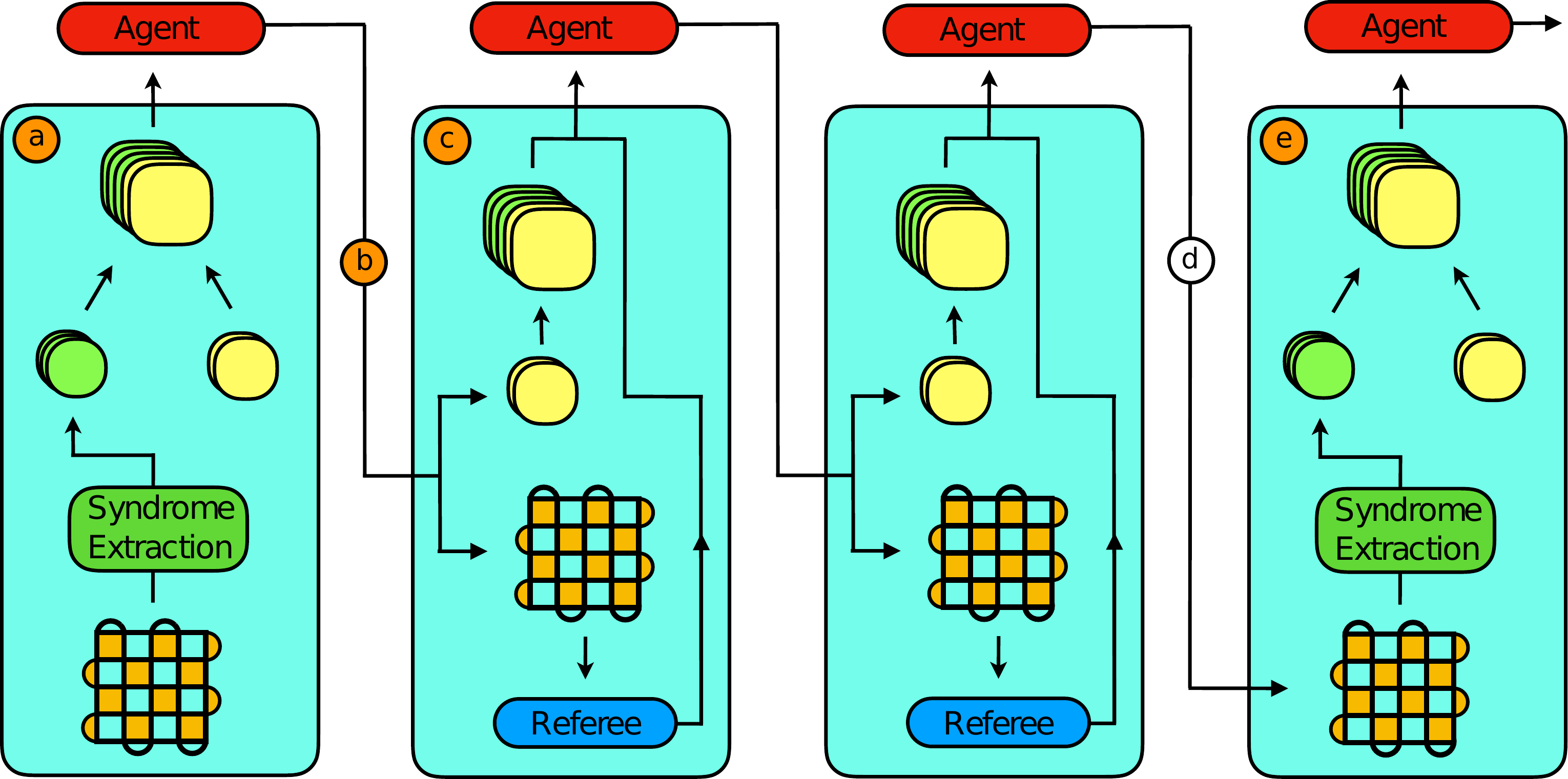}
		\caption{
			An illustration of the various steps occurring within a single episode.
			(a) In an initial step, a faulty syndrome volume is extracted from a state initially in the code space.
			This faulty syndrome volume is combined with an initially empty action history and passed to the agent as the initial state.
			Given an input state, the agent must decide on an action, which it passes to the environment.
			If the action is a Pauli flip which is not already in the action history (b), then the procedure illustrated in box (c) occurs, if the agent requests a new syndrome or repeats an action (d), then the procedure illustrated in box (e) occurs.
			(c) The chosen action is applied to the underlying quantum state.
			From this state the reward for the applied action is determined, while simultaneously a referee decoder decides whether or not the episode is now over - i.e.
			whether the underlying quantum state is now in a terminal state.
			Additionally, the applied action is appended to the action history, which is concatenated with the non-updated syndrome volume and provided to the agent, in conjunction with the reward and terminal state indicator.
			(e) Once again, the underlying quantum state is first updated, and from this updated state both a reward and terminal state indicator are determined (not shown).
			However, the trivial or repeated action of the agent triggers a new round of faulty syndrome measurements.
			After this new round of syndrome extraction, the new syndrome volume is combined with a reset action history and provided to the agent, from which it can once again choose another move.
		}
		\label{f:single_episode}
	\end{figure*}

	We now turn to describing the main topic of this work.
	Namely, we formulate the problem of decoding within the context of \emph{fault-tolerant} quantum computation as a reinforcement learning problem.
	The advantage of this formulation is that it allows for all the methods and techniques of reinforcement learning to be brought to bear on this problem, thereby providing a new toolbox for obtaining diverse \emph{decoding agents}, a novel class of decoding algorithms.
	For clarity, we will utilize the surface code to present all the required elements, however the framework described here could be applied to any stabilizer quantum error correcting code.

	In order to present such a reformulation it is necessary to define the state space $\mathcal{S}$, action space $\mathcal{A}$ and terminal state subset $\mathcal{S}_\mathrm{terminal} \subset \mathcal{S}$, as well as the stochastic process via which the environment generates the tuple $[S_{t+1},R_{t+1},T_{t+1}]$,  when acted upon with action $A_t$.
	As discussed in Section~\ref{s:the_decoding_problem}, within the context of fault-tolerant quantum computation, the goal of the decoding problem is to continuously suppress errors on a logical qubit, to the extent that future logical operations involving this logical qubit can succeed with high probability.
	As such, the fundamental idea is to define all the required elements in such a way that allows for decoding agents to learn to continuously correct errors by performing single qubit Pauli flips on the underlying physical qubits of a given code, remaining ``alive'' as long as future logical operations can succeed with high probability, and being rewarded whenever all errors have been successfully corrected.
	In particular, given some initial logical state $|\psi_0\rangle \in \mathcal{H}_{\mathrm{sc}}$, the goal of the agent is to suppress errors for as long as possible, such that future logical operations can succeed with high probability.

	To provide a framework for achieving this, we consider environments consisting of the following three elements: A \emph{hidden state}, an \emph{error model} and a \emph{referee decoder}.
	At the beginning of any time step $t$, the \emph{hidden state} of the environment $S_{\mathrm{hidden},t}$, will be a list of all the single qubit Pauli flips which have been applied to physical data qubits through the course of the episode, either via errors or as corrections.
	Given both the hidden state $S_{\mathrm{hidden},t}$ and the initial logical state $|\psi_0\rangle$, the current underlying state $|\psi_t\rangle$ of all the physical data qubits could be obtained by applying all the Pauli operations listed in $S_{\mathrm{hidden},t}$ to $|\psi_0\rangle$.
	Note that because we are particularly interested in only the difference between the current state $|\psi_t\rangle$ and the initial logical state $|\psi_0\rangle$, and because we consider only Pauli noise and corrections, we are able to utilise the simple and efficient hidden state description provided here.

	In addition to the hidden state, the \emph{error model} of the environment defines the physical process via which \emph{faulty syndrome volumes} - i.e. a list of violated stabilizers from multiple sequential syndrome measurements - are generated from the hidden state of the environment.
	In Section~\ref{s:results} we utilize the two-stage error model involving separated physical and measurement errors, as described in Fig.~\ref{s:the_decoding_problem}, however we wish to emphasise that in principle a circuit model for syndrome extraction could also be simulated.

	Finally, a \emph{referee decoder} is included to act as a proxy for future logical operations, as such providing a mechanism for the determination of terminal states.
	This referee decoder should be a ``single-shot" decoder which, given a single perfect syndrome, suggests corrections which always move the current state back into the code space, and which may fail by inadvertently suggesting corrections which perform a logical operation.
	In particular, at any stage, the ability of the referee decoder to decode a \emph{perfect} syndrome generated from the current hidden state of the environment will be used as a proxy for the success of future logical operations, and an indicator of whether or not the current state is a terminal state.

	Given these fundamental elements of the environment we define the action space $\mathcal{A}$ to consist of all Pauli $X$ and $Z$ flips on single physical data qubits, along with a special \emph{Request New Syndrome} action.
	Note that Pauli $Y$ flips can be implemented via $X$ and $Z$ flips, and that in practice we imagine all single qubit corrections suggested between successive syndrome measurements would be accumulated, either to be tracked through the computation, or applied simultaneously, as illustrated in Fig.~\ref{agent_decoding}.
	In addition, we define the state space $\mathcal{S}$ to consist of all possible states of the form $S_{t} = \{S_{\mathrm{sv},t},h_t\}$, where $S_{\mathrm{sv},t}$ is a faulty syndrome volume and $h_t$ is the \emph{action history}, a list of all the actions performed by the agent since the syndrome volume $S_{\mathrm{sv},t}$ was generated.

	Finally, we have all the ingredients necessary to describe the rules, illustrated in Fig.~\ref{f:single_episode}, via which the environment generates the tuple $[S_{t+1} = \{S_{\mathrm{sv},t+1},h_{t+1}\},R_{t+1},T_{t+1}]$ when acted upon with action $A_t$.
	In particular, given an environment with hidden state $S_{\mathrm{hidden},t}$, depending on the action $A_t$ one of two procedures, illustrated in Figs.~\ref{f:single_episode} (c) and (e) respectively, will be utilized to generate the tuple $[S_{t+1},R_{t+1},T_{t+1}]$.

	\subsection{Non-Repeated Pauli Flips}\label{s:non-repeated}

	If, as shown in Fig.~\ref{f:single_episode} (c), the action $A_t$ chosen by the agent is any Pauli flip that is not already an element of the action history list $h_t$, then the environment responds as follows:
	\begin{enumerate}
		\item The hidden state of the environment $S_{\mathrm{hidden},t+1}$ is obtained by appending $A_t$ to $S_{\mathrm{hidden},t}$.
		In essence, the agent's chosen correction is applied to the  underlying state $|\psi_t\rangle$, yielding state $|\psi_{t+1}\rangle$.

		\item We set $S_{\mathrm{sv},t+1} = S_{\mathrm{sv},t}$, therefore providing the agent an opportunity to provide more corrections in response to the current syndrome volume.
		In addition, the updated action history $h_{t+1}$ is obtained by appending to $A_t$ to $h_t$.
		The output state is then $S_{t+1} = \{S_{\mathrm{sv},t+1},h_{t+1}\}$.

		\item If $|\psi_{t+1}\rangle$ is equal to the initial logical state $|\psi_0\rangle$ - i.e. if all errors have been corrected at this stage without implementing a logical operation - then $R_{t+1} = 1$, otherwise $R_{t+1} = 0$.
		Technically, this can be determined by checking both that no stabilizers are violated by $|\psi_{t+1}\rangle$, and that $|\psi_{t+1}\rangle$ belongs to the same homology class as $|\psi_0\rangle$.

		\item Finally, the referee decoder is given a perfect syndrome generated from $|\psi_{t+1}\rangle$.
		If the referee decoder can successfully decode this syndrome, then $S_{t+1}$ is not a terminal state and we set $T_{t+1} = 0$.
		If the referee decoder incorrectly decodes the given syndrome, then $T_{t+1} = 0$ and the episode is over.
	\end{enumerate}

	\subsection{Request New Syndrome or Repeated Pauli Flip}

	If on the other hand the agent requests a new syndrome, or chooses an action $A_t \in h_t$ - i.e. an action it has already chosen since first seeing the syndrome volume $S_{\mathrm{sv},t}$ - then the environment responds via the following procedure, as illustrated in Fig.~\ref{f:single_episode} (e).

	\begin{enumerate}
		\item If the action $A_t$ is a Pauli flip already in $h_t$, i.e. if for some reason the agent is choosing to repeat an action, then this chosen correction is applied to the underlying state by appending $A_t$ to $S_{\mathrm{hidden},t}$.
		We denote this new hidden state as $S'_{\mathrm{hidden},t+1}$.
		In this case, the reward $R_{t+1}$ and terminal state indicator $T_{t+1}$ are determined from $S'_{\mathrm{hidden},t+1}$  by the process described in Section~\ref{s:non-repeated} for non-repeated Pauli flips.

		\item If action $A_t$ is the \emph{request new syndrome} action, then the hidden state is not updated, i.e. $S'_{\mathrm{hidden},t+1} = S_{\mathrm{hidden},t}$, and as such $R_{t+1} = R_{t}$ and $T_{t+1} = T_t$.

		\item Now, a new syndrome volume $S_{\mathrm{sv},t+1}$ is generated, via the given error model, from the underlying hidden state $S'_{\mathrm{hidden},t+1}$.
		In the process errors may occur on physical data qubits, and the final hidden state $S_{\mathrm{hidden},t+1}$ is obtained by applying these errors to the intermediate hidden state $S'_{\mathrm{hidden},t+1}$.

		\item Finally, as a new syndrome volume has just been generated, the action history $h_{t+1}$ is reset to an empty list and the total state $S_{t+1} = \{S_{\mathrm{sv},t+1},h_{t+1}\}$ is returned.
	\end{enumerate}

	\begin{figure*}
		\centering
		\includegraphics[width=0.7\textwidth]{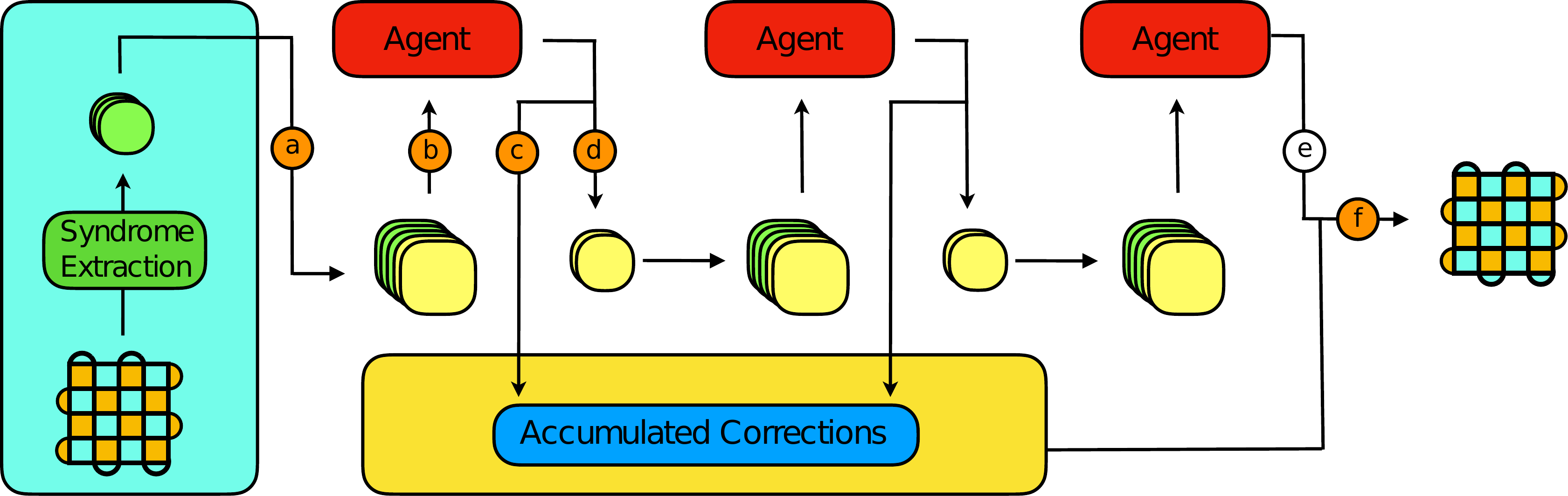}
		\caption{
			The procedure for decoding with a trained agent.
			(a) Given a faulty syndrome volume generated by some experiment, this volume is concatenated with an empty action history to produce a suitable input state for the decoding agent.
			(b) The decoding agent takes in the combined faulty syndrome volume and action history and chooses an action.
			This action is simultaneously (c) added to a list of accumulated corrections and (d) used to update only the action history component of the input state to the agent.
			This updated input state is then given to the agent and procedures (b-d) continue until the agent (e) repeats an action or requests a new syndrome.
			(f) At this stage the corrections could be applied to the lattice state, although in practice they would be tracked through the entire computation.
		}
		\label{agent_decoding}
	\end{figure*}

	Given all the above elements, it is now possible to summarize a complete episode.
	In particular, every episode starts with a resetting of the environment.
	Specifically, as shown in Fig.~\ref{f:single_episode}~(a), given an initial logical state $|\psi_0\rangle \in \mathcal{H}_{\mathrm{sc}}$, represented by a hidden state $S_{\mathrm{hidden},0}$ which is just an empty list, the \emph{request new syndrome} action is applied to the environment to obtain the initial tuple $[S_{1} = \{S_{\mathrm{sv},1},h_{1}\},R_{1},T_{1}]$.
	Now, in any step $t$, given $[S_{t},R_{t},T_{1}]$ the agent needs to decide on an action $A_t$, which is applied to the environment.
	Using the rules described above the environment then generates the tuple $[S_{t+1},R_{t},T_{1}]$, and the process continues until $T_{t+1} = 1$, i.e. until $S_{t+1} \in \mathcal{S}_\mathrm{terminal}$.
	From the construction above one can see that in order to maximise long term discounted cumulative reward, a decoding agent needs to learn how to choose actions that suppress errors, preserving as closely as possible the initial logical state $|\psi_0\rangle$.
	If however, the agent chooses incorrect actions, then errors will accumulate and the referee decoder will no longer be able to correctly decode, indicating the probable failure of subsequent logical gates involving the logical qubit.

	At this stage we have all the ingredients required to obtain decoding agents through the application of a variety of reinforcement learning algorithms.
	In Section~\ref{s:dq_agent} we will present a detailed construction for a deepQ agent, that allows us to apply deepQ learning and obtain decoding agents whose performance is shown in Section~\ref{s:results}.
	Before proceeding however it is worthwhile to emphasize a few points.

	First, it is important to note that the above agent-interaction framework is both code and error model agnostic, provided one has access to a referee decoder, capable of single-shot decoding on perfect syndromes.
	From this perspective, one might view the framework presented here as a tool for leveraging single-shot, perfect-syndrome decoding algorithms, into decoding algorithms for the fully fault-tolerant setting.
	Additionally, as mentioned above, the agent will only accumulate reward provided it can learn to consistently correct any errors that might occur on physical data qubits, without implementing logical operations.
	However, the agent is \textit{not} constrained to return the state back into the code space between successive syndrome measurements.
	In particular, because typical reinforcement learning algorithms take into account not only immediate rewards, but rather discounted cumulative rewards, the agent may learn strategies involving actions whose benefit may not be immediate.
	For example, in the rare event of multiple measurement errors occurring within the extraction of a syndrome volume, creating a highly ambiguous and difficult to decode input state, the agent may choose to only partially decode before requesting a new syndrome volume, in which hopefully less measurement errors will occur and the underlying error configuration may be less ambiguous.
	As such, the decoding agents obtained via this framework may have access to truly novel decoding strategies, that are not currently available to alternative decoding algorithms.

	It is also useful to note that in practice various alternative design choices are possible, and that it is possible to speed-up learning by utilizing various improvements.
	First of all, when choosing exploratory actions, we can restrict the action space to a reduced set of potentially valid corrections.
	In particular, the agent need only consider actions on vertices either involved in a violated stabilizer or adjacent to vertices which have already been acted on.
	This restriction effectively increases the probability of the agent discovering useful actions, and therefore the effectiveness of any exploration phase.
	Second, when generating new syndrome volumes, we discard those cases in which the syndrome is trivial.
	In these cases, the agent does not need to act and hence no useful experience tuple would be generated.
	This allows any experience memory required by the learning algorithm to consist of only useful memories.
	Finally, the structure of the reward mechanism presented here is only one of many possible choices.
	In particular, while this natural reward structure facilitated stable training and the obtaining of decoding agents whose performance is shown in Section~\ref{s:results}, it is an interesting and open question to explore the effectiveness of alternative reward mechanisms.

	Although we have now thoroughly addressed a framework to which reinforcement learning algorithms could be applied to \textit{train} an agent, we have not yet explicitly discussed the implementation of our $q$-function based agent and how it may be used to decode within an experimental setting.
	As illustrated in Fig.~\ref{agent_decoding}, the latter problem is straightforwardly addressed.
	Given a faulty syndrome volume from an experiment, we initialize an empty action history list and combine this with the faulty syndrome volume as an initial input to the decoding agent.
	The trained agent can then be asked directly for an initial correction, which is added to a list of accumulated corrections.
	Simultaneously \textit{only} the action history element of the previous input state is updated.
	This updated input state, containing the original syndrome volume and the subsequently performed corrections, is then given again to the agent as input.
	This process is iterated until the agent either repeats a correction or requests a new syndrome volume.

\section{A deepQ Decoding Agent}\label{s:dq_agent}

	\begin{figure*}
		\centering
		\includegraphics[width=0.9\textwidth]{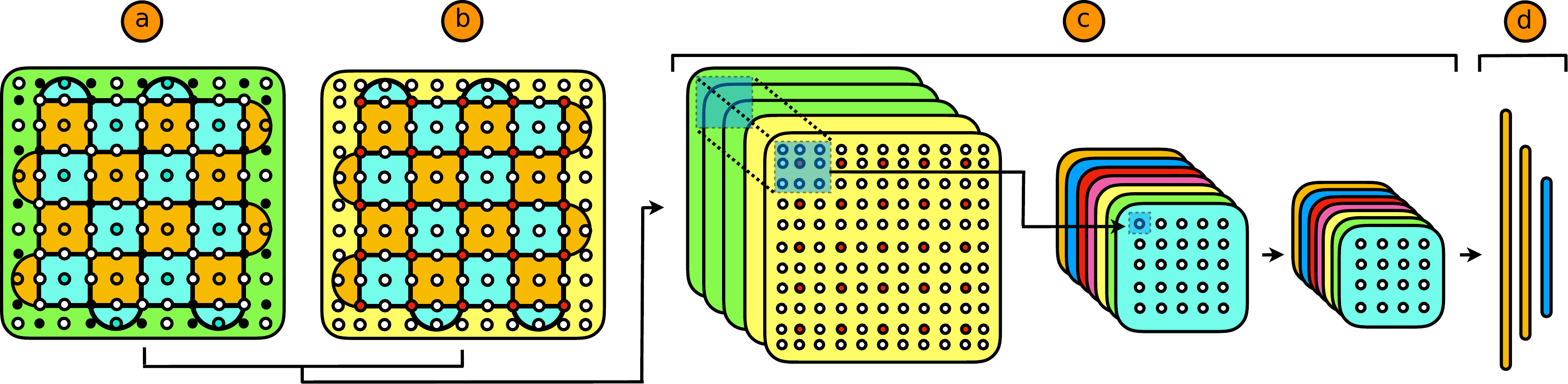}
		\caption{
			Details of a deepQ decoding agent for a $d\times d$ code lattice.
			(a) A single faulty syndrome is embedded into a $(2d +1)\times(2d+1)$ binary matrix, where by default all entries are initially set to $0$.
			Entries indicate by orange (blue) circles are then used to indicate violated $Z$ ($X$) stabilizers on the corresponding plaquette.
			Entries indicated by black circles are set to $+1$ as a way of differentiating different types of stabilizers.
			(b) Using the same embedding, the action history for $Z$ ($X$) flips can be encoded by indicating a previous flip on a specific vertex qubit via the entry corresponding to that vertex.
			(c) By stacking the syndrome and action history slices, the total state $S_t$ can be constructed in a form suitable for input to a convolutional neural network.
			(d) To complete the deepQ network, we stack a feed forward neural network on top of the convolutional layers.
			In particular, the final layer of this network has $|\mathcal{A}|$ activations, each of which encodes $q(S_t,a)$ for an action $a$.
		}
		\label{f:agent}
	\end{figure*}

	As discussed in the previous section, many different reinforcement learning algorithms could now be applied within the framework presented here.
	However, in order to provide a concrete example and proof-of-principle, we will specialize to deepQ learning, which has been previously utilized to obtain agents capable of human-level control in domains such as Atari~\cite{RLMnih15}.
	As mentioned briefly in Section~\ref{s:reinforcement_learning}, a variety of now standard tricks are required to get deepQ learning to work in practice, and we will not present these details here, referring the reader to the relevant references~\cite{RLMnih15,RLvan2016deep,RLschaul2015prioritized,RLwang2015dueling} or associated code repository~\cite{DeepQDecoding}.
	However, there are various details concerning the construction of the deepQ agent which may be useful for applying alternative deep reinforcement learning algorithms within this framework, and as such we will present these details in this section.

	In particular, in Section~\ref{s:decoding_as_rl} we described how at the beginning of any time step $t$ the agent is supplied with the state $S_{t}=\{S_{\mathrm{sv},{t}},h_{t}\}$, where $S_{\mathrm{sv},{t}}$ is a faulty syndrome volume, given as a list of violated stabilizers from successive syndrome measurements, and $h_{t}$ is the action history list.
	In deepQ learning, and in many other deep reinforcement learning algorithms, this state $S_t$ needs to be provided as the input to a deep neural network.
	For example, in the case of deepQ learning, the state $S_t$ is the input to the deepQ network parametrizing the $q$-function from which the agent is partially deriving its policy.
	As such, utilizing an encoding of $S_t$, which allows for the use of appropriate neural networks, is important.

	In Fig.~\ref{f:agent} we have illustrated the encoding of $S_t$ which was used to facilitate the use of deep convolutional neural networks to parametrize the $q$-function.
	In particular, as shown in Fig.~\ref{f:agent} (a) and (b), we can embed a $d\times d$ code lattice into a $(2d + 1)\times(2d+1)$ binary matrix, where each entry corresponds to either a plaquette, a vertex, or an edge of the lattice.
	As shown in Fig.~\ref{f:agent} (a), we can then use a single such binary matrix to encode each of the faulty syndromes, by using the entries corresponding to plaquettes to indicate violated stabilizers, and the remaining entries to differentiate both blue and orange, and bulk and boundary plaquettes.
	Similarly, as illustrated in  Fig.~\ref{f:agent} (b) we can use two such binary matrices to encode the action history, by using one matrix to indicate the physical data qubits on which $X$ flips have already been applied, and the other binary matrix to indicate the data qubits on which $Z$ flips have already been applied.
	As can be seen in Fig.~\ref{f:agent} (c), the total state $S_t$ can then be obtained by stacking the action history slices on top of the faulty syndrome slices, effectively creating a multi-channel image suitable as input for a convolutional neural network.
	In particular, the strength of this encoding is that any convolutional filter, such as the one indicated in Fig.~\ref{f:agent} (c), then isolates all relevant information from some local patch of the lattice - in particular, the violated stabilizers and their type, as well as the previously applied actions.
	Finally, the deepQ network we utilize is completed by stacking a feed forward neural network on top of multiple convolutional layers.
	In particular, the final layer of this network has $|\mathcal{A}|$ activations, each of which encodes $q(S_t,a)$ for an action $a$.

\section{Results}\label{s:results}

	As a demonstration, we have 
	utilized deepQ learning within the framework presented in Section~\ref{s:decoding_as_rl}, to obtain deepQ decoding agents for both bit-flip and depolarizing noise, with faulty syndromes, for a $d=5$ surface code lattice.
	All the code used to obtain these agents is supplied in the corresponding \textit{DeepQ-Decoding} code repository~\cite{DeepQDecoding}, and as such we will provide only an overview of the details here.
	In particular, for a single fixed set of hyper-parameters, we have exploited 
	the original deepQ algorithm~\cite{RLMnih15} with annealed $\epsilon$-greedy exploration, implemented via the keras-rl library~\cite{plappert2016kerasrl} and incorporating doubleQ updates~\cite{RLvan2016deep} and a dueling network architecture~\cite{RLwang2015dueling}.
	In addition, we have made use of a custom training procedure, described in Appendix~\ref{A:training}, for sequentially iterating through increasing error rates, while simultaneously performing a hyper-parameter grid search at each error-rate iteration.
	For both error models we utilized a convolutional deepQ network, as illustrated in Fig.~\ref{f:agent}, consisting of three convolutional layers, followed by a single feed-forward layer before the final output layer.
	Specifically, if we describe a single convolutional layer with a three-tuple $[n,w,s]$, where $n$ is the number of filters, $w$ is the filter width and $s$ is the stride, and a single feed forward layer via a two-tuple $[n,d]$, where $n$ is the number of neurons and $d$ is the output drop-out rate, then from input to output our deepQ networks had the base structure,
	\begin{equation}
		[[64,3,2],~[32,2,1],~[32,2,1],~[512,0.2],~[|\mathcal{A}|, 0]],
	\end{equation}
	where $|\mathcal{A}|$ is the size of the action space.
	Additional minor modifications as required for dueling methodologies were also included~\cite{RLwang2015dueling}.
	All other additional hyper-parameters used to obtain each decoding agent, along with histories of each training procedure, are provided in Appendix~\ref{A:parameters}.

	\begin{figure*}
		\centering
		\includegraphics[width=\linewidth]{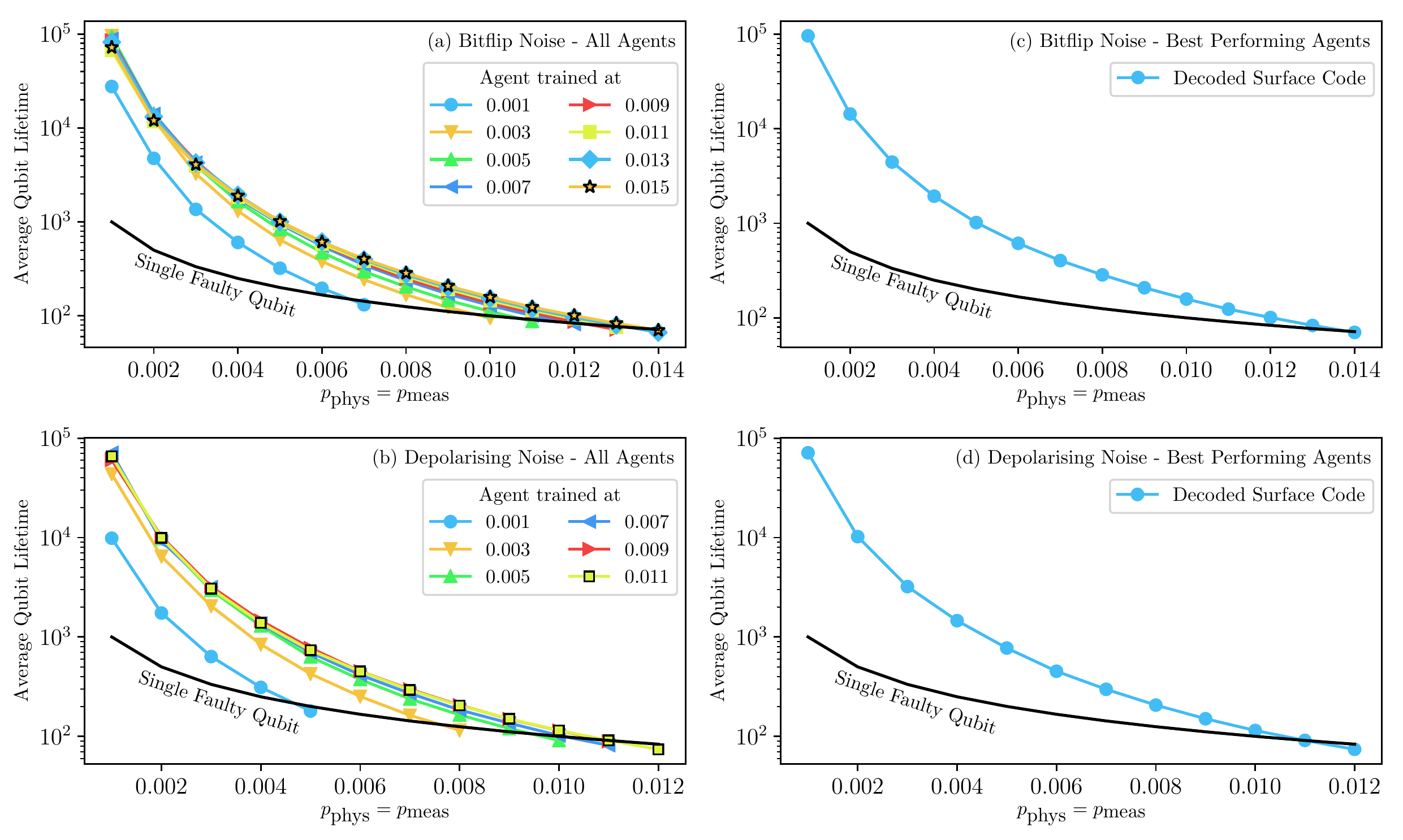}
		\caption{
			(a,b) Performance of all agents obtained during the iterative training procedure.
			Each agent was evaluated at increasing error rates, until the average lifetime of the logical qubit actively decoded by the agent became less than the average lifetime of a single faulty qubit.
			(c,d) Results obtained by using the best performing decoding agent for each error rate.
		}
		\label{f:results}
	\end{figure*}

	We considered both bit-flip and depolarizing noise models.
	For both of these error models we considered the measurement of a single syndrome to consist of two separate error channels, as illustrated in Fig.~\ref{f:decoding_problem}.
	For bit-flip noise, in the first error channel - the physical error channel - a Pauli $X$ flip was applied to each physical data qubit with probability $p_{\mathrm{phys}}$.
	For depolarising noise, the physical error channel acted by applying to each physical data qubit, with probability $p_{\mathrm{phys}}$, either a Pauli $X$, $Y$ or $Z$ flip, with equal probability.
	For both noise models, after the physical error channel, the true syndrome was calculated, after which the second measurement error channel was applied, in which the value of each stabilizer measurement was flipped with probability $p_{\mathrm{meas}}$.
	For all simulations we set $p \defeq p_{\mathrm{phys}} = p_{\mathrm{meas}}$, and used a syndrome volume depth of 5.
	Also, note that for bit-flip noise, as only $X$ corrections were necessary, we had $|\mathcal{A}| = d^2 + 1$ - i.e. the agent could perform either the \textit{request new syndrome} action or a single qubit $X$ flip on any individual physical data qubit.
	For depolarising noise, we restricted the agent to only $X$ and $Z$ flips (as $Y$ errors can be corrected via both an $X$ flip and a $Z$ flip), such that $|\mathcal{A}| = 2d^2 + 1$ rather than $|\mathcal{A}| = 3d^2 + 1$.

	To evaluate the performance of our trained decoders we used the procedure described in Fig.~\ref{agent_decoding}, where the agent selected actions via the final $q$-function in a purely greedy manner, with repeated actions or the request of a new syndrome triggering new syndrome volumes.
	As referee decoders, we utilized fast feed-forward neural network based homology class predictors, trained in a supervised manner, as per Refs.~\cite{Torlai10, Varsamopoulos17}.
	All utilized referee decoders are included in the DeepQ repository~\cite{DeepQDecoding}.
	In particular, the referee decoder was used to check after every action of the agent whether or not a terminal state had been reached, and the length of a single episode was reported as the number of individual syndromes seen by the agent (i.e. the number of times the two-fold error channel was applied) before a terminal state was reached.
	For each error rate, the average lifetime of the actively decoded logical qubit was determined by the average episode length, over a number of episodes that guaranteed at least $10^6$ syndromes were seen by the decoding agent.
	This average logical qubit lifetime should be compared to the average lifetime of a single faulty qubit.

	The final results are then shown in Fig.~\ref{f:results}.
	In particular, Fig.~\ref{f:results} (a, b) shows the performance of all decoding agents obtained during the iterative training procedure (as described in Appendix~\ref{A:training}), while Fig.~\ref{f:results} (c, d) shows the results obtained by the using the best performing decoding agent for each error rate.
	For bit-flip (depolarising) noise we find that for approximately $p < 1.3\times 10^{-2}$ $(p < 1.1\times 10^{-2})$ there is a decoding agent for which the average lifetime of the actively decoded $d=5$ logical qubit is longer than the average lifetime of a single faulty qubit.
	While these results should not be interpreted as rigorous thresholds, they can be seen as proof-of-principle demonstrations of the feasibility of using decoding agents, and in particular deepQ agents, as decoders for the fully fault-tolerant setting.
	In particular, in the context of near term fault-tolerant demonstrations of quantum computation, we insist that 
	what is most important is the life-time of the logical qubit at the specific error-rate given by the experimental device.
	In this respect, our proof-of-principle decoders perform competitively at code distances and error-rates expected in near-term devices.

	Furthermore, there are various points worth emphasising.
	First, both the neural network architecture and reinforcement learning algorithm used here are comparatively simple with respect to the current state-of-the-art~\cite{RLmnih2016asynchronous,RLSilver17b,RLsilver2017mastering,RLSilver2016}, and were chosen to allow for the execution of the required training procedure with the available computational resources.
	As such, it is expected that utilization of either more sophisticated neural network architectures or learning algorithms, coupled with the computational resources required for implementing the required training procedures, could allow one to obtain significantly better results.
	Furthermore, as discussed in Ref.~\cite{chamberland2018deep}, using dedicated hardware it is expected that the forward pass time of various applicable neural network architectures (including the architecture used here) can be brought below the time-scales necessary for near-term experiments.
	Factoring in the ability to straightforwardly apply the techniques here to circuit-level noise renders decoding agents promising candidates for near-term fault-tolerant experiments.

\section{Conclusion}\label{s:conclusions}

	We have shown that the problem of decoding within the setting of fault-tolerant quantum computation can be naturally reformulated as a reinforcement learning problem.
	In particular, we have provided an agent-environment framework which allows for the application of diverse reinforcement learning algorithms, and provided a proof-of-principle demonstration by training deepQ decoding agents for the $d=5$ surface code, for both bit-flip and depolarizing noise with faulty syndrome measurements.
	It is important to stress that this framework is both code and error model agnostic, and hence can be directly used to train decoding agents, a novel class of flexible decoding algorithms, for a wide variety of other experimentally relevant settings.
	Additionally, the recent use of more sophisticated reinforcement learning techniques and neural network architectures to demonstrate super-human performance in complex domains such as Chess and Go~\cite{RLSilver17b,RLsilver2017mastering,RLSilver2016}, strongly indicate that the initial results presented here could be successfully improved and extended upon (given appropriate computational resources).
	With the development of dedicated special purpose hardware for the fast implementation of neural networks, decoding agents could indeed provide a practical solution for decoding in near-term fully fault-tolerant quantum computation.

	There is a plethora of natural and interesting ways of extending these initial results.
	First and foremost it would be of interest to explore the performance of alternative decoding agents, obtained via different reinforcement learning algorithms, for a wider class of codes.
	Crucially, a key feature of this approach is the ability to straight-forwardly tackle general error-models, even including models exhibiting intricate correlations.
	Equally important is the consideration of techniques allowing for the scaling of neural network decoders to larger code distances.
	In this work we have focused on the decoding of idling logical qubits during quantum computation, encoded via surface code patches of fixed size.
	For such a setting recent literature provides neural network architecture suggestions for scaling to larger code distances~\cite{Ni18}, although an alternative method for future research could be found in the use of multiple communicating decoding agents, simultaneously decoding on overlapping sub-lattices of the total code.

	However, current surface code based approaches to large scale fault-tolerant quantum computing 
	require methods for decoding of irregularly shaped and constantly changing surface code patches, 
	formed via lattice surgery and code deformation~\cite{Litinski18b,Fowler18}.
	Hence, to provide a truly complete toolbox for fault-tolerant quantum computing it is necessary to provide 
	decoding algorithms which do not explicitly depend on the underlying code lattice.
	For the case of decoding agents, this would require the development of methods circumventing the requirement to retrain agents for different shape surface code patches.
	The aforementioned approach of multiple communicating agents acting on fixed size sub-lattices could however provide a promising approach to this problem.

	Furthermore, the framework presented here relies on the availability of a referee decoder, with access to the state of the underlying physical data qubits.
	As a result, the framework given here can not be used to train decoding agents only on experimental data, and in practice one would be required to first estimate an error model for the experimental setup.
	In order to remove this restriction it would be of interest to investigate modifications to the framework given here, which require only experimentally generated syndromes.
	One such modification would be to consider ``single-shot'' episodes, which terminate as soon as the agent has returned the current state into the code space.

	Finally, due to the inherent design of the procedure via which these decoding agents are obtained, in which discounted future performance is valued over immediate rewards, these agents have the potential to learn decoding strategies not available to alternative decoding algorithms.
	It may therefore be insightful to construct test-cases (i.e.\ specific error volumes) to infer the learned strategy, and use this information in the design of model-tailored decoding algorithms. 
	It is the hope that the present work constitutes a significant step forward in the understanding of the applicability of notions of 
	reinforcement learning -- with its ability to predict sophisticated situations that require an understanding of the impact of present actions into the future -- in quantum information science and in the quantum technologies.

	\begin{acknowledgments}
		The authors gratefully acknowledge helpful and insightful discussions with Daniel Litinski, Nicolas Delfosse, Aleksander Kubica, 
		Thomas Jochym-O'Connor, Paul Baireuther, James Wootton and Hendrik Poulsen Nautrup. Two of us (J.E. and E.vN) would like to thank Roger Melko, Titus Melko and Simon Trebst for the invitation to the workshop on "Machine Learning for Quantum Many-body Physics" in Dresden in June 2018, where two ongoing research programs have been merged into the present collaboration.
		Additionally, the authors would like to thank J\"{o}rg Behrmann for incredible technical support, without which this work would not have been possible.
		RS\ acknowledges the financial support of the Alexander von Humboldt foundation.
		MSK\ is supported by the DFG (CRC183, project B02).
		EvN\ is supported by the Swiss National Science Foundation through grant P2EZP2-172185.
		JE\ is supported by DFG (CRC 183, EI 519/14-1, and EI 519/7-1), the ERC (TAQ), the Templeton Foundation, and the BMBF (Q.com).
		This work has also received funding from the European Union's Horizon 2020 research and innovation programme under grant agreement No 817482 (PASQUANS).
	\end{acknowledgments}

\appendix

\section{Distributed Iterative Training with Simultaneous Hyper-Parameter Optimization}\label{A:training}

\begin{figure}
		\centering
		\includegraphics[width=\linewidth]{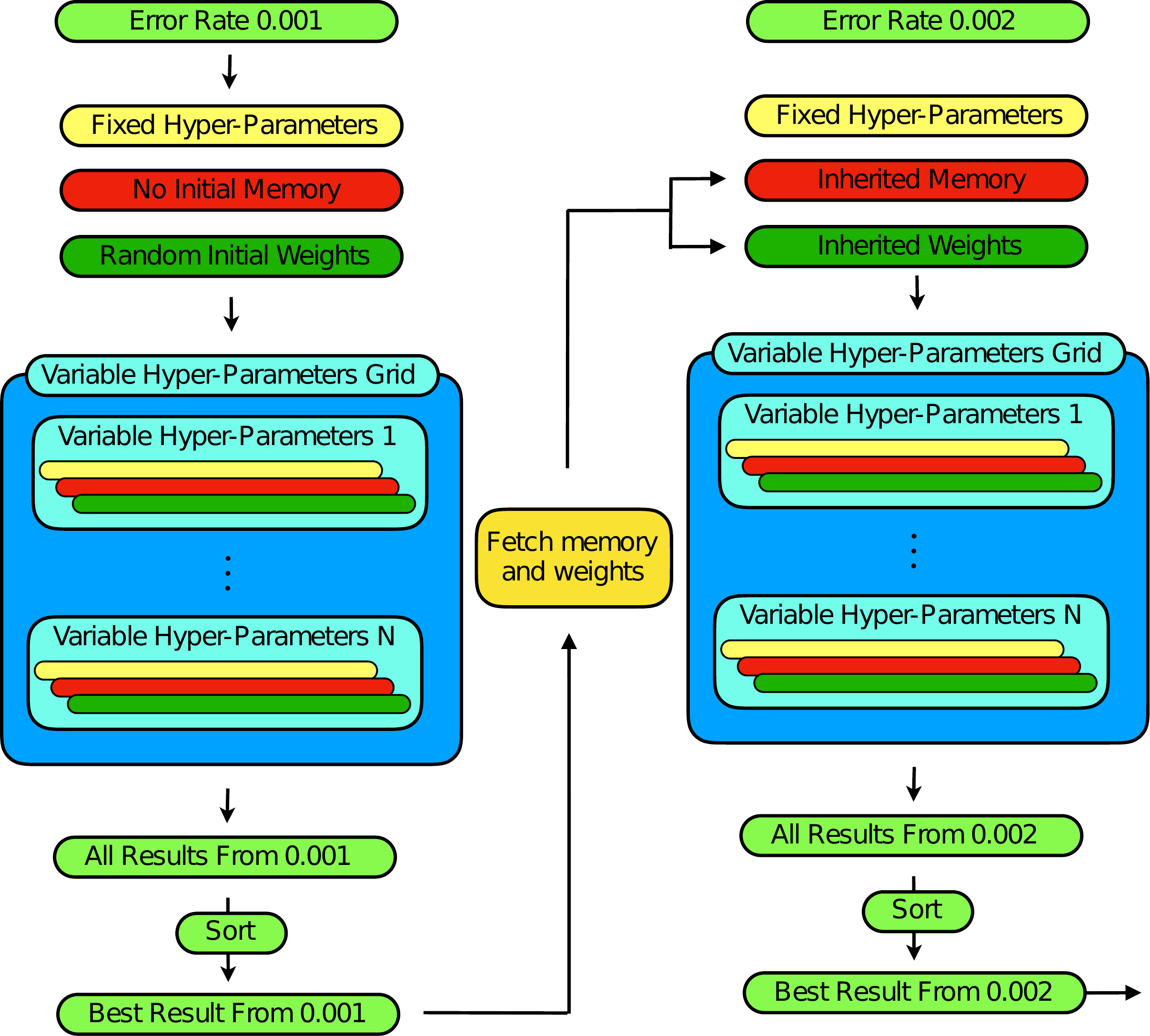}
		\caption{
			Iterative training procedure, via a hyper-parameter grid search at each error rate.
			After training multiple decoding agents at a given error rate, one for each point in the specified hyper-parameter grid, the results are sorted.
			The weights and experience memory from the optimal decoding agent are then used as the initial weights and experience memory for a new round of training procedures, again one for each point in the hyper-parameter grid, but at an increased error rate.
		}
		\label{f:iterative_training}
	\end{figure}

	In order to obtain optimal decoders for multiple error rates we implemented a custom iterated training procedure, involving a hyper-parameter grid search at each error rate, as illustrated in Fig.~\ref{f:iterative_training}.
	All code for implementing this procedure on an HPC cluster can be found in the associated \emph{DeepQ-Decoding} repository~\cite{DeepQDecoding}.
	In particular, this procedure involves the following steps:

	\begin{enumerate}
		\item Choose an initial error rate $p_1$, and fix the values for any hyper-parameters which should remain constant throughout the entire training procedure.
		\item For all hyper-parameters not yet fixed, specify a list of values to be used for training - i.e.\ specify a hyper-parameter grid over which one would like to search for an optimal hyper-parameter configuration.
		\item Train multiple decoding agents at the initial error rate, one for each hyper-parameter configuration in the specified grid.
		Each agent is initialized with an empty experience memory and random initial neural network weights.
		The training of all agents can be done simultaneously in a distributed manner.
		\item Given the results from all agents trained at $p_1$, sort the results and store the neural network weights and experience memory from the optimal decoding agent.
		\item Increase the error rate to $p_2$.
		Once again, train multiple decoding agents, one for each hyper-parameter configuration in the specified grid.
		However, this time all agents are initialized with the experience memory and neural network weights from the optimal agent at $p_1$.
		\item Iterate this procedure until a specified final error rate $p_n$, or until the performance of the optimal agent is worse than that of a single faulty qubit.
	\end{enumerate}

\section{Agent Hyper-Parameters and Learning Curves}\label{A:parameters}

\begin{figure}
		\centering
		\includegraphics[width=\linewidth]{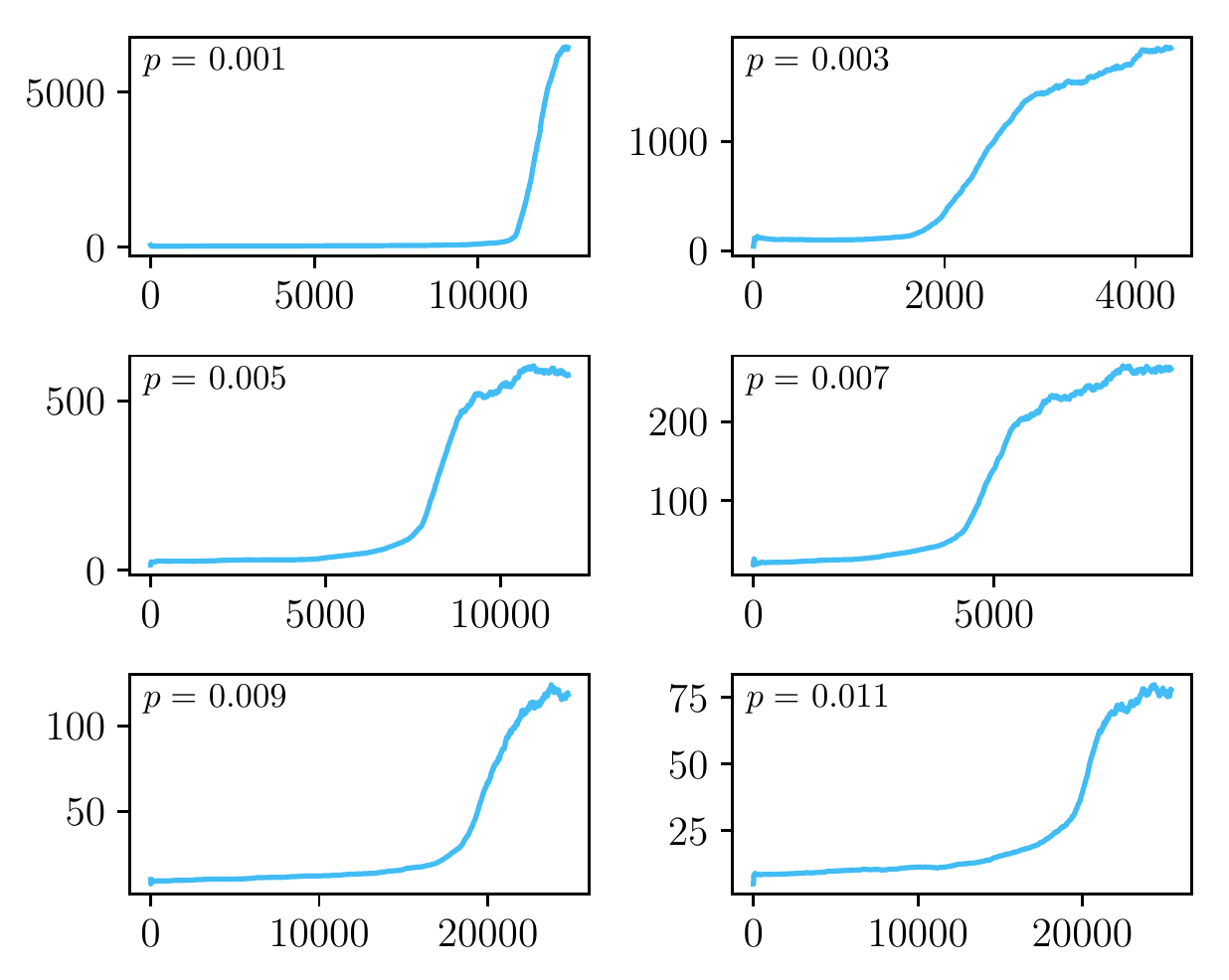}
		\caption{
			Training histories for optimal depolarising noise agents.
			The $x$-axis shows the number of episodes, and the $y$-axis shows the rolling average of the decoded qubit lifetime.
			Hyper-parameters for each agent are given in Table~\ref{t:agent_hyper_parameters}.
		}
		\label{f:training_results_dp}
	\end{figure}

\begin{figure}
		\centering{}
		\includegraphics[width=\linewidth]{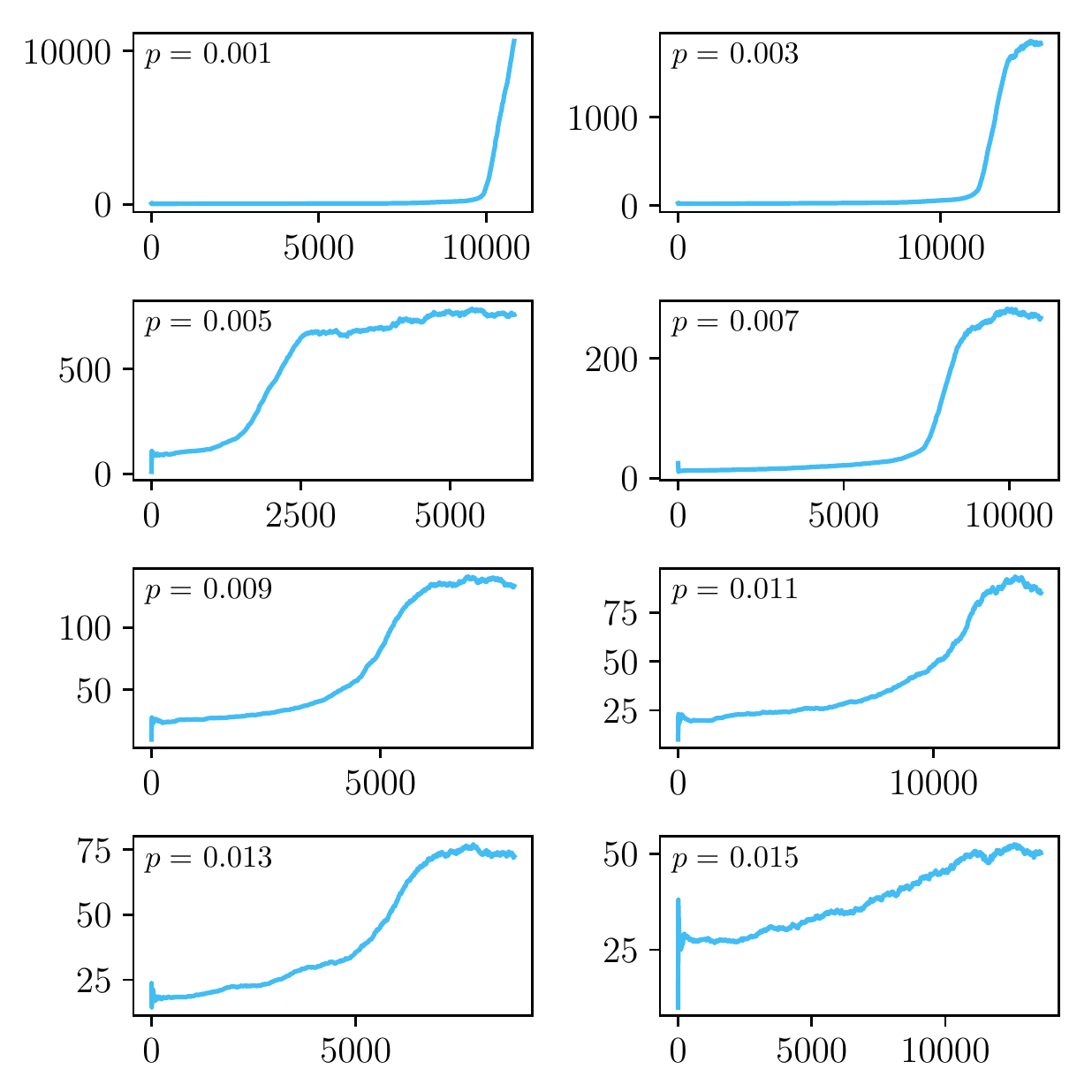}
		\caption{
			Training histories for optimal bit-flip noise agents.
			The $x$-axis shows the number of episodes, and the $y$-axis shows the rolling average of the decoded qubit lifetime.
			Hyper-parameters for each agent are given in Table~\ref{t:agent_hyper_parameters}.
		}
		\label{f:training_results_x}
	\end{figure}

	We implemented the distributed iterative training procedure described in Appendix~\ref{A:training}, for both bit-flip and depolarising noise (all utilized code can be found in the \emph{DeepQ-Decoding} repository~\cite{DeepQDecoding}).
	In particular, the initial error rate was set to $p_1 = p_{\mathrm{phys}} = p_{\mathrm{meas}} = 1\times10^{-3}$, and incremented by $2\times10^{-3}$ in each iteration.
	As described in Section~\ref{s:results}, from input to output the neural network architecture was as follows:
	\begin{equation}
		[[64,3,2],~[32,2,1],~[32,2,1],~[512,0.2],~[|\mathcal{A}|, 0]],
	\end{equation}
	with additional modifications required for dueling methodologies~\cite{RLwang2015dueling} implemented automatically by keras-rl~\cite{plappert2016kerasrl}.
	Single convolutional layers have been described with a three-tuple $[n,w,s]$, where $n$ is the number of filters, $w$ is the filter width and $s$ is the stride, and single feed forward layers via a two-tuple $[n,d]$, where $n$ is the number of neurons and $d$ is the output drop-out rate.
	All agents were trained via the original deepQ algorithm~\cite{RLMnih15}, implemented via the keras-rl library~\cite{plappert2016kerasrl}, with annealed $\epsilon$-greedy exploration, doubleQ updates~\cite{RLvan2016deep} and dueling networks~\cite{RLvan2016deep}.
	Both the fixed hyper-parameters and variable hyper-parameter grids utilized during the training procedure are specified in Table.~\ref{t:hyper_parameters}.
	Figs.~\ref{f:training_results_dp} and~\ref{f:training_results_x} then show the training history of the optimal decoding agent at each error rate, while the associated values for the variable hyper-parameters are given in Table~\ref{t:agent_hyper_parameters}.

	\begin{table}[hb]
	\begin{center}
		\begin{tabular}{| c | c | }
			\hline
			\multicolumn{2}{|c|}{Fixed Hyper-Parameters} \\
			\hline
			Batch size & $32$  \\ 
			Rolling average length & $1\times10^3$ \\ 
			Stopping patience (in episodes) & $1\times10^3$  \\
			Maximum training steps & $1\times10^6$  \\ 
			Memory buffer size & $5\times 10^4$ \\ 
			Syndrome volume depth & $5$ \\
			Discount factor $\gamma$ & 0.99 \\
			\hline
			\multicolumn{2}{|c|}{Variable Hyper-Parameters} \\
			\hline
			Initial $\epsilon$ & $\{1, 0.5, 0.25\}$ \\ 
			Final $\epsilon$ & $\{0.04, 0.02, 0.001\}$  \\
			Number of exploration steps & $\{1,2\}\times10^5$  \\ 
			Learning rate (LR)& $\{10,5,1,0.5\}\times10^{-5}$  \\ 
			Target network update frequency & $\{2500, 5000\}$ \\ 
			\hline
		\end{tabular}
		\caption{
			Training Hyper-Parameters.
		}
		\label{t:hyper_parameters}
	\end{center}
	\end{table}

	\begin{table}[hb]
	\begin{center}
		\begin{tabular}{| c | c | }
			\hline
			\multicolumn{2}{|c|}{Bitflip Noise} \\
			\hline
			0.001 & $[2\times10^5,1,0.02,1\times10^{-5},5000]$  \\ 
			0.003 & $[2\times10^5,1,0.02,1\times10^{-5},5000]$  \\ 
			0.005 & $[1\times10^5,0.25,0.001,1\times10^{-5},2500]$  \\ 
			0.007 & $[1\times10^5,1,0.02,1\times10^{-5},5000]$  \\ 
			0.009 & $[1\times10^5,0.5,0.04,1\times10^{-5},5000]$  \\ 
			0.011 & $[2\times10^5,0.5,0.04,1\times10^{-5},2500]$  \\ 
			0.013 & $[1\times10^5,0.5,0.001,1\times10^{-5},5000]$  \\ 
			0.015 & $[2\times10^5,0.25,0.04,1\times10^{-5},2500]$  \\ 
			\hline
			\multicolumn{2}{|c|}{Depolarising Noise} \\
			\hline
			0.001 & $[2\times10^5,1,0.001,5\times10^{-5},5000]$  \\ 
			0.003 & $[1\times10^5,0.25,0.02,1\times10^{-5},5000]$  \\ 
			0.005 & $[2\times10^5,0.5,0.001,1\times10^{-5},5000]$  \\ 
			0.007 & $[1\times10^5,1,0.02,1\times10^{-5},5000]$  \\ 
			0.009 & $[2\times10^5,1,0.02,1\times10^{-5},2500]$  \\ 
			0.011 & $[2\times10^5,1,0.02,5\times10^{-6},2500]$  \\ 
			\hline
		\end{tabular}
		\caption{
			Hyper-parameters for optimal agents at each error rate, in the form [\textit{number of exploration steps, initial} $\epsilon$ \textit{, final} $\epsilon$\textit{, learning rate, target network update frequency}]
		}
		\label{t:agent_hyper_parameters}
	\end{center}
	\end{table}

	\clearpage
	\bibliography{RL_Decoding_for_FTQCNotes.bbl}

\end{document}